# Observation of spontaneous magnetization jump and field-induced irreversibility in $Nd_5Ge_3$

BIBEKANANDA MAJI[1], K. G. SURESH[1,*] AND A. K. NIGAM[2]
[1]Magnetic Materials Laboratory, Department Of Physics,
Indian Institute Of Technology Bombay, Mumbai- 400076, India
[2]Tata Institute Of Fundamental Research,
Homi Bhabha Road, Mumbai- 400005, India

*Abstract*

We report the observation of spontaneous and ultra-sharp jumps in the low temperature magnetization isotherms of polycrystalline $Nd_5Ge_3$. Field-induced and ultra-sharp jumps are also seen in resistivity and heat capacity data. These jumps are accompanied by field-induced irreversibilities. The consistency seen in these three data clearly shows that the spin, electronic and lattice states are strongly coupled. Time-induced growth of the ferromagnetic phase is observed at a constant field and temperature, implying the metastability of the magnetic phase in the low field region. Various experimental findings point towards a strong field-induced magneto-structural irreversibility in this compound.

*Corresponding author (email: suresh@phy.iitb.ac.in, Fax: +91-22-25723480)

## 1. Introduction

Recently, compounds exhibiting field-induced metamagnetic phase transition in which the system undergoes a transition from a low magnetic moment state to a high magnetic moment state have attracted a lot of attention [1]-[4]. The metamagnetic transitions in some compounds are found to be first order in nature and the presence of strong magneto-structural coupling gives rise to many interesting magnetic and related properties. Examples are certain phase-separated colossal magnetoresistive manganites [5], a few intermetallics like $Gd_5Ge_4$ [6, 7], doped $CeFe_2$ [8, 9] and full Heusler alloys [10, 11]. The martensitic strains resulting from the structural change or at least a distortion at the metamagnetic transition are found to control the magnetic and the related properties in some of these materials. One of the most striking features of the martensitic scenario is that the transition from the low field antiferromagnetic (AFM) state to the field-driven ferromagnetic (FM) state occurs via ultra-sharp steps at low temperature, even in polycrystalline samples. Recently, Hardy et al.[12] have shown that in $Pr_{0.5}Ca_{0.5}Mn_{1-x}Ga_xO_3$ manganite, the steps can be found in the time evolution of magnetization i.e., in a situation where both the temperature and magnetic field are kept constants. A similar observation was also reported in manganite thin films by Wu and Mitchell [13]. Spontaneous magnetization and resistivity steps were seen in $(La_{0.5}Nd_{0.5})_{1.2}Sr_{1.8}Mn_2O_7$ by Liao et al [14]. Strong protocol dependence of magnetic properties is one of the most important signatures of the martensitic scenario [12, 14].

In a recent paper, a field-induced and irreversible AFM to FM transition was reported in $Nd_5Ge_3$ single crystals when the field was applied along the *c*-axis [15]. In fact this compound has been known for many years and detailed neutron diffraction study is available for a long time [16]. According to this study, $Nd_5Ge_3$ crystallizes in the $Mn_5Si_3$- type hexagonal structure with the space group $P6_3/mcm$. The Nd atoms occupy two crystallographic inequivalent sites, namely 6(g) and 4(d), whereas the Ge atoms occupy 6(g) site. Based on the neutron diffraction study, it was also suggested that there are two types of antiferromagnetic arrangements in this compound. The Nd atoms at 6(g) have a collinear antiferromagnetic arrangement, whereas the ones at 4(d) site form a canted antiferromagnetic structure making an angle of 31º with the *c* axis. Using the magnetization data, the same authors concluded that there is a spin-flop transition giving rise to a large hysteresis at low temperatures.

As part of our investigations on metamagnetic systems, we have chosen polycrystalline $Nd_5Ge_3$ and subjected it for a detailed magnetization (*M*), electrical resistivity (*ρ*) and heat capacity ($C_P$) study. The aim of this work was to investigate the dynamics of the metamagnetic transition, by collecting these data under various experimental protocols. We observe field-induced step-like magnetic transition in *M*(*H*), *ρ*(*H*) as well as $C_P$(*H*) curves almost at the same field. Furthermore, a field-induced irreversibility has been observed in all these three properties. Another interesting observation is a step-like spontaneous growth of magnetization as a function of time, at the critical field. This kind of a jump, where both the temperature and magnetic field are kept constant, is observed in very limited number of compounds. It is even more significant in the light of the fact that observation of such a jump in an intermetallic compound is rather unique and therefore, we probe this phenomenon in detail in this paper.

2. **Experimental Details**

The polycrystalline $Nd_5Ge_3$ was prepared by arc melting a stoichiometric mixture of the constituent elements of Gd (99.9 - at. % purity) and Ge (99.999-at. % purity), in a water-cooled copper hearth, in high purity argon atmosphere. The resulting ingot was turned upside down and remelted four times to ensure homogeneity. The weight loss after the final melting was less than 0.5 %. The structural analysis of the sample was performed by collecting the room temperature powder x-ray diffractogram (XRD) using Cu-K$_α$ radiation. The magnetization measurements were carried out using a vibrating sample magnetometer attached to a Physical Property Measurement System (Quantum Design, Design, PPMS-6500). The heat capacity (using the relaxation technique) and the electrical resistivity (using a linear four probe method) measurements were carried out in PPMS.

## 3. Experimental Results

Rietveld refinement of the XRD pattern of $Nd_5Ge_3$ at room temperature shows that there are no detectable impurities in the compound. The crystal structure is found to be $Mn_5Si_3$- type hexagonal with the space group $P6_3/mcm$. Lattice parameters, $a=b=8.7508(3)$ $\overset{o}{A}$ and $c= 6.6345(2)$ $\overset{o}{A}$, obtained from the refinement are in good agreement with the literature [15, 16].

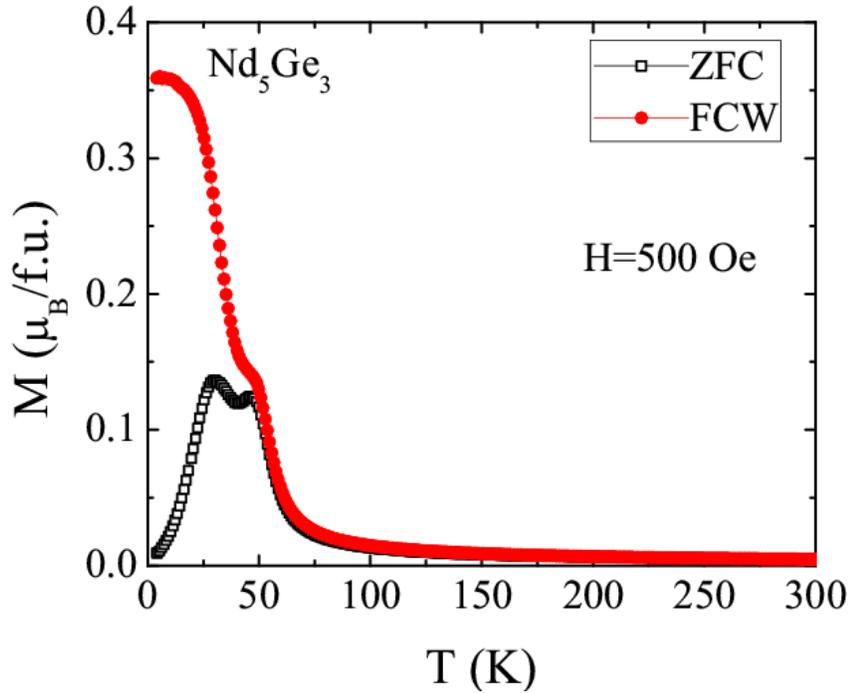

**Fig. 1**: Temperature dependence of ZFC and FCW magnetization of $Nd_5Ge_3$ in a field of 500 Oe.

Fig. 1 shows the temperature dependence of dc magnetization in the presence of a field ($H$) of 500 Oe both under the zero field cooled (ZFC) and the field cooled warming (FCW) conditions. In both these cases, the data was collected during the warming cycle. In the FCW mode, both the cooling and the measuring fields were 500 Oe. We see that the ZFC curve exhibits two peaks at $T_N = 46$ K and $T_t = 30$ K. This indicates the presence of two types of antiferromagnetic states in $Nd_5Ge_3$. One can see that the anomaly at 46 K is present even in the FCW curve, whereas the peak at 30 K almost disappears. It suggests that the AFM configuration responsible for $T_t$ anomaly is easily destroyed and a ferromagnetic state is achieved by the application of field [16]. The ZFC and FCW magnetization curves exhibit a large thermomagnetic irreversibility below $T_N$. This indicates the possibility of coexistence of antiferromagnetic/spin glass state/magnetic glass, at low temperatures.

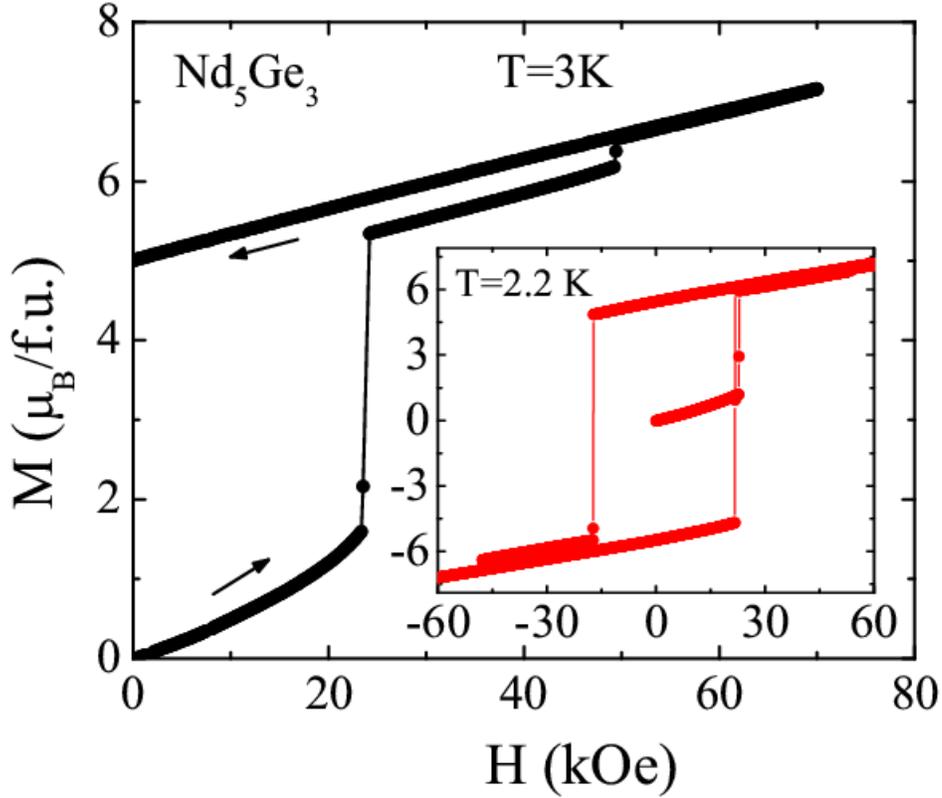

**Fig. 2**: Magnetization isotherm at 3 K in Nd$_5$Ge$_3$. The inset shows the 5-loop *M-H* plot at 2.2 K upto a field of 60 kOe.

In fig. 2, we show the isothermal *M(H)* plot at *T*=3 K. The measurement has been performed by sweeping the field at a rate of 190 Oe/s, after zero field cooling the sample from temperatures well within the paramagnetic region to 3 K. The isothermal *M* (*H*) displays two field-induced metamagnetic transitions, at critical fields of 23 kOe and 49 kOe. During the first metamagnetic jump the magnetic moment of the alloy changes from 1.6 to 5.3 $\mu_B/f.u.$ and during the second jump it increases from 6.2 to 6.5 $\mu_B/f.u.$. In the reverse cycle the magnetization curve does not display any step and gives rise to a large remanence. This indicates that the field-induced transformation is completely irreversible. Such an irreversibility is usually not seen in the metamagnetic transitions in antiferromagnets. The inset of fig. 3 shows the 5 loop hysteresis plot at 2.2 K. As can be seen, the remanent magnetization is almost retained upto a field of about -25 kOe. The symmetry between the positive and negative field cycles and the rectangular shape of the hysteresis loop are quite striking. Another noteworthy observation is that the virgin curve (1$^{st}$ path) lies just outside the envelope curve (5$^{th}$ path), indicating the presence of supercooling [8]. This gives an indication that the field-induced transition is of first order in nature. However, the separation between the virgin curve and the envelope curve in Nd$_5$Ge$_3$ is not as prominent as in other materials showing supercooling/phase coexistence such as doped CeFe$_2$ [3, 8].

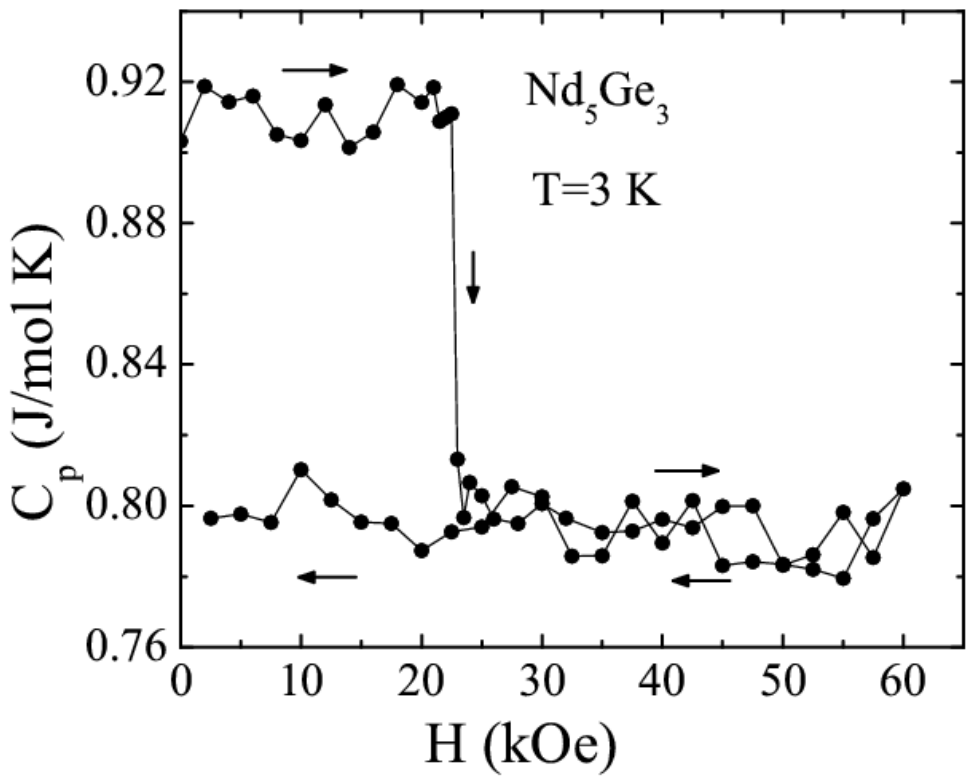

**Fig. 3**: Heat capacity as a function of applied field in Nd$_5$Ge$_3$ at 3 K.

Fig. 3 shows the variation of the heat capacity as a function of field at 3 K. $C_P$-$H$ data shows a large decrease at about the same field where the first magnetization step is observed. This decrease is extremely sharp with a width of about 900 Oe. The second metamagnetic transition is not seen in the $C_P(H)$ plot, probably due to the data fluctuation. The occurrence of the step in the heat capacity at nearly the same magnetic field as that of the magnetization step underlines the fact that the magnetic and electronic transitions are strongly coupled in this compound. Another very interesting feature seen in the $C_P$-$H$ plot is that on the decreasing field path, the heat capacity value almost remains a constant down to the zero field point. This field-induced irreversibility is quite a remarkable observation and is in accord with the irreversibility seen in the magnetization data.

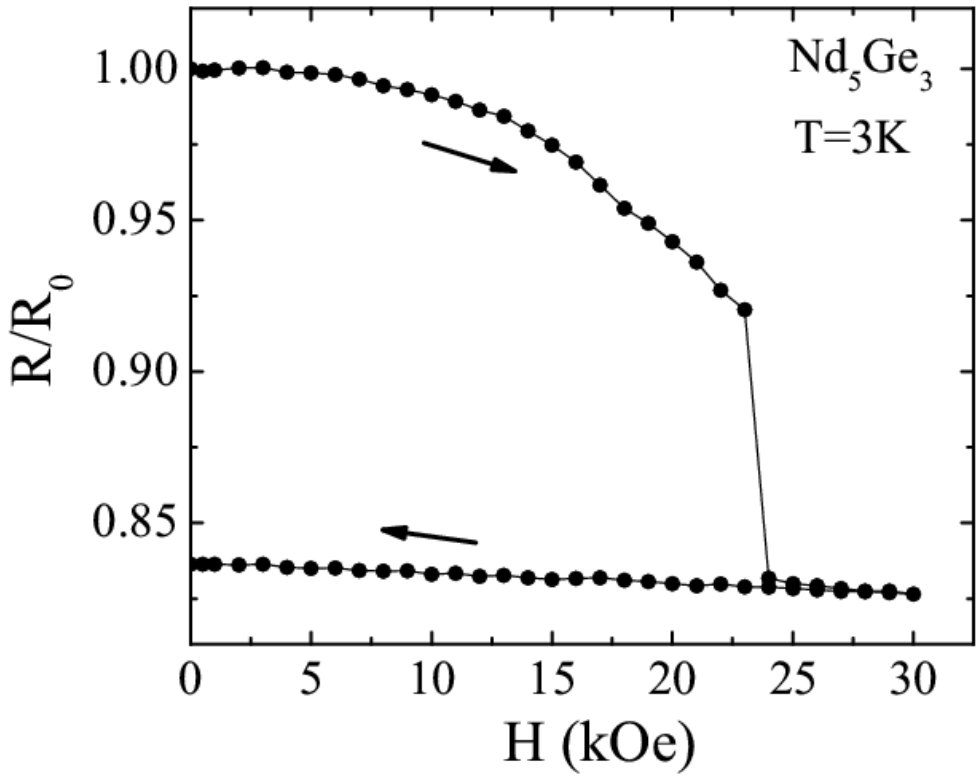

**Fig. 4**: Normalized electrical resistance as a function of field at 3 K in $Nd_5Ge_3$.

In order to further probe the change in the magnetic state at the critical field, we have measured the electrical resistivity as a function of field at 3 K. Fig. 4 shows the variation of the resistance ($R$) normalized to the zero field value, as a function of applied field. Interesting, the *R(H)* also shows a sudden decrease at about 23 kOe. The field-induced irreversibility seen in the magnetization and heat capacity is also seen in the resistivity isotherms. Therefore, the consistency seen in the three different measurements is quite striking. The magnetoresistance is found to be about 16% at 3 K for a field of about 25 kOe.

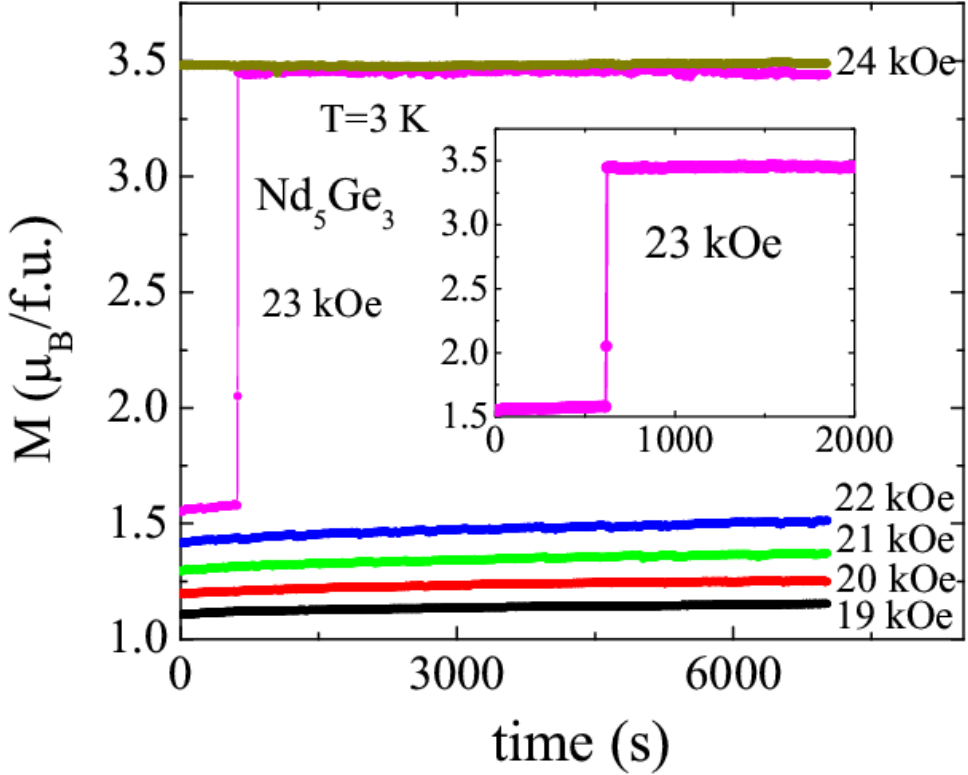

**Fig. 5**: Variation of magnetization in $Nd_5Ge_3$ as a function of time, under different fields. The inset shows the transition region at 23 kOe in the expanded scale.

In view of the metastability of the magnetic state at low fields, we have further probed the time dependence of the magnetization to study the relaxation effect. The magnetic relaxation experiment is a very useful and effective tool to study the details of the dynamics of the magnetization jumps [13, 14]. Therefore, we performed the magnetic relaxation experiment at a fixed temperature (3 K) and in different fields in the vicinity of the critical field corresponding to the first metamagnetic transition observed in the isothermal *M(H)* curve. The sample was cooled in zero field from room temperature down to 3 K. After the temperature got stabilized, a field of 19 kOe was applied and the magnetization was measured in this field for a time period of 7000 s. The magnetization was about 1.1 $\mu_B/f.u.$ throughout the waiting time. The experiment was repeated many times after following the same protocol and applying higher fields, in steps of 1 kOe. As can be seen from Fig. 5, for fields upto 22 kOe, the magnetization behavior is almost identical to that at 19 kOe, except for a small and gradual increase in the value. However, when the field was increased to 23 kOe, we see an ultra-sharp jump in the magnetization (as shown in expanded fig. in the inset). The reproducibility of this remarkable feature has been checked by repeating the experiment for at least two different pieces of virgin sample. It is seen that the magnetization jump takes place over a very short time interval of about 10 s. When the experiment was repeated in a field of 24 kOe, the magnetization started from a value which is almost equal to the post-jump value at 23 kOe and remains constant throughout the duration of 7000 s. Therefore, it is clear that the transition that has occurred at 23 kOe is irreversible. The time spent by the system at a fixed magnetic field and temperature before the start of the magnetization jump is called incubation time, which is found to be about 610 seconds in the present case. For the higher field ($H > 23$ kOe), the relaxation curve is almost flat, without any considerable time dependence. But the system shows a weak relaxation when the field is lower than 23 kOe. It was found that the *M(t)* curves (normalized to the value at *t*=0) for these latter fields can be fitted to the equation $M(H,t) = M(H,0) + [M(H,\infty) - M(H,0)]\{1 - \exp[-t/\tau(H)]\}$ where *M(H,t)* represents the magnetization at time *t* and field *H*, $\tau(H)$ is the relaxation time for the given field. This relaxation time is related to the magnitude of the energy barrier between two metastable states [14]. The obtained $\tau$ values vary from 2300 to 3200 s. The corresponding values in the case of some phase separated manganites exhibiting similar features lie in the range of 5600-7100 s [12].

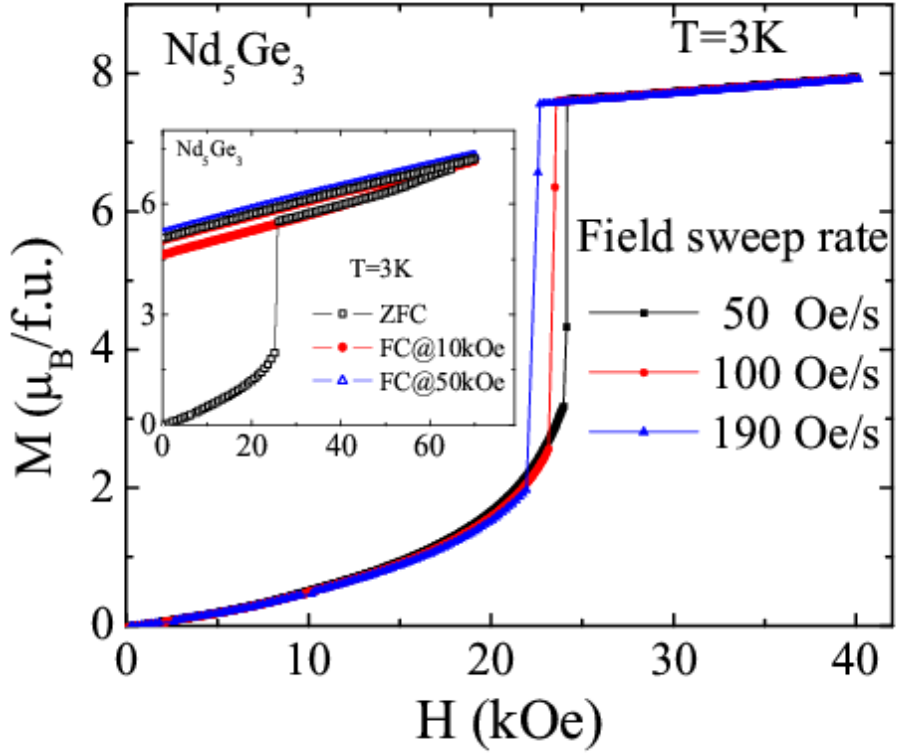

**Fig. 6**: Field sweep rate dependence of magnetization isotherm in Nd$_5$Ge$_3$ at 3 K. The inset shows the *M-H* plot at 3 K, after field cooling in different fields.

In view of the anomalous features seen in the data presented above, we have further investigated the magnetization dynamics by studying the magnetic field sweep rate and cooling field dependencies. Fig. 6 shows the *M(H)* loop at 3 K recorded for three different field sweep rates. Before collecting the data, every time the sample was zero field cooled to 3 K from 200 K. We see that the critical field at which the magnetic step arises is strongly dependent on the field sweep rate. It is quite evident from the fig. that the critical field ($H_C$) decreases with increasing field sweep rate. The $H_c$ value is ~24 kOe when the sweep rate is 50 Oe/s, which decreases to 23 kOe and 22 kOe when the sweep rate is enhanced to 100 Oe/s and 190 Oe/s respectively. This feature of field sweep rate dependence points towards the role of martensitic strains in determining the magnetization variation of the system. For a lower sweep rate the lattice gets more time to adapt to the martensitic strains, resulting in the upward shift of the critical field [14].

Effect of field cooling on the *M* (*H*) isotherm was also studied for this compound. For this measurement, the sample was cooled from 200 K to 3 K in the presence of different fields. After stabilizing 3 K, the cooling field was reduced to zero and then the magnetization was recorded subsequently by increasing the field up to 70 kOe. The data was also taken for the decreasing field cycle. This variation, along with the ZFC data (for comparison) is shown in the inset of Fig. 6. As can be seen, there is no jump in the magnetization (as seen in the ZFC mode) when the sample is field cooled even in a nominal field of 10 kOe. On the other hand, the low-field magnetization is found to be large under field cooling. There is a slight increase in the magnetization with increase in the cooling field. This implies that the FM fraction increases with cooling field. Similar behavior is observed in some phase-separated manganites [17] and doped $CeFe_2$ [8, 18].

## 4. Discussion

The experimental results presented above are very similar to those seen in many phase separated oxides in which martensitic scenario is known to influence the magnetic properties. There are some similarities with a few intermetallic compounds as well. It is quite clear that the features seen here are not merely magnetic in origin, but has contribution from strong magneto-structural coupling. Though several explanations have been proposed by different authors for the step-like metamagnetic transitions in different classes of materials, the most prominent one in the present case seems to be the martensitic effect. The magnetic state of $Nd_5Ge_3$ at low temperatures seems to be predominantly antiferromagnetic. But, with the application of a suitable field, the FM phase grows both as a function of field as well as time. In this framework, all the features shown in the previous section can be attributed to the field-induced lattice distortion (martensitic like transition) in this compound. A strong lattice distortion by an external magnetic field via magneto-elastic coupling seems to be the reason behind the martensitic effect in this case. It is to be noted that Doerr et al. [19] have reported an irreversible magnetovolume effect through magnetostriction studies in $Nd_5Ge_3$. The field-induced FM state is a crystallographically distorted phase in this case. Due to the difference in the crystallographic structures between the low and high field phases, an elastic strain develops at the FM/AFM interface. When a field is applied, the FM phase is likely to grow but the interfacial (martensitic) strain acts against this to stop the growth of the FM phase. As the field is continuously increased, the driving force acting on the moments increases. When the applied field is large enough to overcome the elastic strain energy, the FM phase grows in a catastrophic manner, which results in sudden jumps in magnetization. Along this avalanche process, the magnetic energy decreases while the elastic energy increases, a balance which can lead the system to be frozen in another metastable state. The overall transition may thus proceed by successive jumps between metastable states [20] and this will influence all the related properties. However, the noteworthy difference that we see in the present case as compared to many other systems with martensitic features is the strong field-induced irreversibility that is seen in magnetization, resitsivity and heat capacity isotherms. For example, though there are many similarities between doped $CeFe_2$ and $Nd_5Ge_3$, there is no field-induced irreversibility (either in magnetization or in resistivity) in the former. Furthermore, the effect of supercooling in the present case is not as prominent as in the case of doped $CeFe_2$. On the other hand, we find that there are many striking similarities between $Nd_5Ge_3$ and the Co substituted NiMnSb full Heusler alloy [21]. A strong field-induced irreversibility in NiCoMnSb as revealed by magnetization, heat capacity and resistivity data, as in the case of $Nd_5Ge_3$ has been reported recently. The consistency seen in the magnetic, transport and heat capacity data clearly shows that the spin, electronic and lattice degrees of freedoms are strongly coupled. It should also be mentioned here that the magnetization and heat capacity jumps reported by Ghivelder et al.[22] in phase separated $LaPrCaMnO_3$ are also identical to those seen in $Nd_5Ge_3$. However, these authors have attributed the temperature rise associated with the magnetocaloric effect for the jumps. Dho and Hur [23] have reported quite a similar resistivity behavior in $PrCaMnO_3$. It is to be highlighted that most of the materials reported with exactly similar results (as in $Nd_5Ge_3$) belong to the family of magnetic oxides. In this context, it is also worthwhile to mention that in the intermetallic compound $Nd_7Rh_3$, though the field-induced irreversibility was evident in the resistivity data, such a signature was absent in the magnetization isotherm [24]. These authors attributed a percolative conduction mechanism for the irreversibility in the resistivity data. Moreover, the spontaneous magnetization jump is also not reported in this case. Therefore, it is quite clear that the scenario present in $Nd_5Ge_3$ is fundamentally different from that of $Nd_7Rh_3$. We feel that as far as spontaneous magnetization jump is concerned, $Nd_5Ge_3$ is the only rare earth intermetallic compound that shows features similar to the phase separated oxides.

It is quite evident that the field-induced irreversibility evident from different measurements in the present case is due to some structural disorder, which traps the field-driven, high magnetization/low resistance/low heat capacity FM state. This gives rise to the arrest of the field-driven FM state, giving rise to the large irreversibility. It is of interest to note that even though this is a stoichiometric compound, defects seem to be present and play a decisive role in determining the magnetic and other properties, especially at low temperatures. At this point, we do not rule out the role of quenched disorder in arresting the FM phase.

## 5. Conclusion

We have shown that the stoichiometric intermetallic compound $Nd_5Ge_3$ shows many anomalies in the magnetization, resistivity and heat capacity data. Ultra-sharp jumps are seen in the low temperature isotherms of all these three data. The most important anomaly is the field-induced irreversibility in all these three properties, which suggests that the spin, electronic and lattices states are strongly coupled in this compound. The memory effect seen here is quite remarkable and consistent in various measurements. Another striking observation is the evolution of the FM phase as a function of time, at a fixed temperature and at the critical field. Sweep rate and cooling field dependencies point towards the role of martensitic strains in the anomalous behavior. The features seen in this compound indicate the presence of disorder in arresting the field-driven FM phase, thereby giving rise to the irreversibility. A very recent report on $Gd_5Ge_4$ reveals the role of magnetic deflagration phenomenon in determining the anomalous martensitic-like properties [7]. Taking into account the similarities between $Gd_5Ge_4$ and $Nd_5Ge_3$, we feel that that a similar study will be useful to shed more light on the anomalous properties of this compound.

**Acknowledgment**

The authors thank D. Buddhikot for his help in resistivity measurements.